\documentclass{moriond}
\newcommand{\Photo}{}

\usepackage{physics}

\def\be{\begin{equation}}
\def\ee{\end{equation}}
\def\bea{\begin{eqnarray}}
\def\eea{\end{eqnarray}}

\begin{document}
\vspace*{4cm}
\title{Tests of the quantisation of gravity with gravitational wave detectors}
\author{ G. Tobar}

\address{Department of Physics, Stockholm University,\\
SE-106 91 Stockholm, Sweden}

\maketitle

\abstracts{The quantisation of gravity remains a central open question in fundamental physics, with direct experimental evidence still elusive. Here, we revisit the problem of single graviton detection by proposing a novel approach: using gravitational wave detectors with particle-number-resolving capabilities. Unlike traditional schemes relying on microscopic systems and spontaneous processes, our method leverages macroscopic quantum resonators subject to stimulated transitions from gravitational waves. These gravitational-wave-induced energy transitions mirror the key features of the photoelectric effect and can serve as detectable signatures of individual graviton absorption events. We demonstrate that such a detector, sensitive to quantised energy exchanges, would provide strong evidence for the quantisation of the gravitational field under the assumption of energy conservation at the level of individual transitions. Furthermore, we outline a concrete path toward experimental realisation using high-Q, cryogenically cooled, multi-mode resonant mass detectors with optomechanical phonon counting, as well as through measuring side-band scattered photons in interferometric detectors. }

\section{Introduction}
The quantisation of gravity remains one of the most important unresolved problems in fundamental physics. While quantum field theory provides the most successful framework for the other fundamental forces, with both theoretical consistency and experimental verification, gravity has long remained an outlier, both in terms of its theoretical treatment as well as concrete experimental verification of its quantum nature. Tests for signatures in the linearised regime have been proposed with quantum superpositions of source masses \cite{bose2017spin,marletto2017gravitationally} or noise in gravitational wave detectors \cite{parikh2020noise,PhysRevLett.127.081602,Guerreiro2022quantumsignaturesin}. Despite these approaches, a particularly well-known challenge is the direct detection of single gravitons, which would constitute the strongest evidence to date for the quantum nature of gravity. Dyson famously argued such detection is impossible \cite{DYSONFREEMAN2013IAGD}.

Here, we show that gravitational wave (GW) detectors with particle-number-resolving readout can act as single graviton counters. These detectors can register individual quantum transitions triggered by the absorption of single gravitons from gravitational radiation—thus providing the first evidence of the quantisation of the gravitational field in analogy with the photo-electric effect. We argue that this is feasibly, despite tests of quantum statistics of gravitational radiation still remaining far out of experimental reach \cite{carney2023}.

\section{Revisiting graviton detection}

The problem of single graviton detection has been considered previously \cite{DYSONFREEMAN2013IAGD}, with well known results – it is fundamentally impossible with any realistic detector. The reason for this is because conventional analyses assume spontaneous graviton emission or absorption by atomic systems, with rates as small as $\Gamma \sim 10^{-40} \; \mathrm{Hz}$ for atomic detectors absorbing or emitting single gravitons \cite{Boughn_2006}. 

However, these estimates rely on two central assumptions:
\begin{itemize}
\item Gravitons interact individually with atomic-scale systems of microscopic spatial extent.
\item The relevant processes are spontaneous, rather than stimulated by a gravitational field containing a macroscopic number of quanta \footnote{Dyson did, however, consider certain stimulated processes—specifically, gravito-electric transitions in atomic detectors interacting with a thermal graviton flux from the Sun. He concluded that even such classical states of gravitational radiation would be incapable of inducing transitions analogous to the photoelectric effect.}.
\end{itemize}
These assumptions underpin prior arguments asserting the infeasibility of single-graviton detection. Microscopic systems, such as atoms, exhibit extremely weak coupling to gravitational radiation. Moreover, spontaneous processes are inherently further suppressed due to the negligible strength of the interaction of the quantum vacuum of linearised gravitational interaction with matter.

In contrast to proposals that rely on preparing and detecting individual gravitons with atomic systems - an approach often deemed impractical due to extremely weak gravitational interactions, we propose a more accessible route: using quantum transitions in macroscopic quantum systems induced by gravitational waves (GWs) as effective indicators of single graviton exchange. This approach does not require a single-graviton input state (or coupling to the vacuum), nor is it constrained by the exceedingly small interaction cross-sections typical of atomic-scale systems. We draw inspiration from the historical development of quantum theory, particularly the photoelectric effect. There, the detection of individual photons through electron emission provided pivotal evidence for the quantisation of the electromagnetic field. Similarly, we propose a gravito-phononic analogue: gravitational waves interacting resonantly with mechanical resonators can trigger discrete, measurable transitions from the ground state to the first excited state of an acoustic oscillator (with the phonon taking the role of the electron in the photo-electric effect). These transitions would correspond to single graviton absorption events, and therefore serve as a graviton detector – a detector that clicks when a single graviton is absorbed. Such gravito-phononic-detection schemes, especially when implemented with resonant mass detectors capable of readout in the phonon number basis, offer a promising pathway to probe the quantum nature of gravity, in analogy with the photo-electric effect for photons. 

\section{Gravitationally-Induced-Quantum-Jumps as photo-electric analogues}
Recent work in Nature Communications 15, 7229 (2024), has shown that a gravitational wave can induce quantum jumps between the energy levels of a massive quantum acoustic resonator. This establishes the feasibility of observing the gravito-phononic effect (the gravitational version of the photo-electric effect). The transition probability scales as:
\begin{equation}
    P_{0 \rightleftharpoons 1} \approx \frac{L^2}{4 \pi^4} \frac{M}{\omega \hbar} h_0^2 \nu^4 t^2 \mathrm{sinc}^2\left(\frac{\omega-\nu}{2} t\right),   
\end{equation}
which mirrors the threshold frequency behaviour observed in the photo-electric effect. These results imply that a macroscopic enhancement to graviton-matter interactions can be achieved using resonators with large mass and quality factor Q, such as resonant mass detectors.

In contrast to spontaneous processes, stimulated transitions can occur with appreciable rates. For instance, an $1800 \; \mathrm{kg}$ Aluminum bar detector subject to a GW strain amplitude of $h_0~\sim~5~\times 10^{-22}$ (comparable to GW150914) can exhibit a transition rate $\Gamma_{\mathrm{stim}}~\sim~1~\;~\mathrm{Hz}$. To make the gravitational photo-electric analogy more precise, we consider projective energy measurements on the bar’s energy levels, allowing for the resolution of discrete phonon excitations analogous to photo-electron emission. We emphasize three key signatures that are analogous to photo-electric emission that manifest in gravito-phononic transitions:

\begin{enumerate}
    \item Threshold frequency: transition probability sharply increases above a specific detector resonance frequency.
    \item Amplitude-independence: the energy of ejected phonons does not depend on the amplitude of the GW.
    \item Near instantaneous production of photo-electrons: Non-zero probability for gravito-phonon production at arbitrarily small times.
\end{enumerate}
Such signatures replicate all core signatures of the photo-electric effect, with the linearised gravitational field in place of the electromagnetic field, and the gravitational wave detector in place of the atom.

\section{Semi-Classical vs Quantum Interpretations}
These features can be partially modelled with semi-classical dynamics, where the gravitational field remains classical while the detector is quantised. However, semi-classical models fail to conserve energy at the level of individual quantum transitions \cite{PhysRevD.9.853,TOBAR2025139785}. Thus, under the assumption of energy conservation, the observed discrete energy exchanges imply that the gravitational field itself must be quantised, and that the energy of gravitational waves is partitioned into discrete quanta $\hbar \omega$.

This conclusion aligns with the original logic of the photo-electric effect \cite{TobarGermain2024Dsgw}: although the first experiments did not produce single-photon-sources \cite{shenderov2024}, the discrete energy exchange provided very strong evidence for the existence of photons – as the alternative semi-classical explanations violate energy conservation (as outlined in Ref.~\cite{TOBAR2025139785}, non-energy conserving transitions are statistically suppressed in full QED, while in the semi-classical theory energy conservation is violated by $\hbar \omega$ for transitions between detector eigenstates), which is restored by the use of full QED (or the quantisation of the linearised gravitational field in the gravito-phononic-case). In this sense, constructing such a graviton-counter would provide the most substantial evidence for the quantisation of gravity to date.

\section{Progress towards experimental realisation}
\subsection{Resonant Mass Detectors}

We propose a practical architecture using multi-mode resonant bars, in which chirping gravitational waves (e.g. from neutron star mergers) couple to the normal modes of a macroscopic oscillator. The resonantly absorbed energy can be read out optomechanically from the end mass of the bar, which can in-principle be nano-scale \cite{TobarGermain2025Dkgf}, enabling the graviton-to-phonon conversion process to be confirmed through established optomechanical read-out techniques. In order to implement the graviton detection scheme proposed here, the following requirements need to be satisfied:
\begin{itemize}
    \item The resonator needs to be sufficiently cooled to enable the resolution of single graviton absorption events above thermal noise.
    \item The modern phonon counting techniques of Ref.~\cite{Tobar_2023_CSL} need to be extended to multi-mode resonant mass detectors as suggested in Ref.~\cite{TobarGermain2025Dkgf}.
    \item The detector geometry, mass and material needs to be developed to ensure a resonance frequency in the interferometric detection band, with sufficiently high $Q \sim 10^9$.
\end{itemize}
While technically challenging, the resonant mass detector proposal was significant in that it opened a first practical pathway to graviton detection from known sources of gravitational radiation. However, the stringent Q-factor and thermal noise requirements, particularly given the limitations imposed by the proposed materials, may severely constrain its experimental feasibility. This motivates a refocusing of experimental efforts on the interferometric approach - based on filtering single-sideband scattered photons from interferometric detectors \cite{6dzy-4c45} that will offer a more experimentally viable route to graviton detection.

\subsection{Interferometric Detectors}
Some initial estimates can provide some input on the possible feasibility of this, for example, if we follow Ref.~\cite{PhysRevX.15.011034} (and later apply it to the LIGO configuration), it analyses how for general interferometers, length fluctuations, where $\delta L_\mathrm{12}$ is the difference in length of arm 1 and the length of arm 2 of the interferometer. This occurs with constant of proportionality, 
\begin{equation}
    G = \pdv{\Dot{N}}{\langle \delta L_\mathrm{12}^2  \rangle} = \frac{k P_{BS}}{\hbar c},
\end{equation}
this constant is known as the optical gain factor of the interferometer. Now, we can relate this photon flux to the differential arm length fluctuations, as,
\begin{equation}
    \Dot{N} = G \langle \delta L_\mathrm{12}^2  \rangle.
\end{equation}
Now integrating over some finite bandwidth $\epsilon$, 
\begin{equation}
\mathcal{S}_{\dot{N}}^\phi(\epsilon)=G \frac{S_L^\phi(f)}{2},
\end{equation}
where for the LIGO configuration we can extract the relevant constant of proportionalities as, $G_L = \frac{\bar{\alpha}^2 \omega_0^2}{4 \kappa L^2}$ and $G_h = \frac{\bar{\alpha}^2 \omega_0^2}{4 \kappa}$. Importantly, it can be used for classical noise sources to convert between noise in the measured length of the interferometer and output photon flux noise. If we now perform a quick insertion of typical LIGO detector parameters as in Ref.~\cite{6dzy-4c45}, we find that typical Gain factors are $G_L = 5.25 \times 10^{39}$, $G_h = 8.40 \times 10^{46}$. Importantly, this allows for the translation of the LIGO noise curves (specifically the classical stochastic noise sources) as
\begin{equation}
    S_{\dot{N}}^{\mathrm{cl}}(f)=\frac{5.62 \times 10^{47}}{1+(f / 400)^2}\frac{S_h^{\mathrm{cl}}(f)}{2} ,  
\end{equation}
having used optical cavity decay rates of $\frac{\kappa}{2\pi} = 400 \; \mathrm{Hz}$ for the LIGO cavity, with $f$ being the displacement noise frequency. Converting typical strain curves using $G_h$, say examining some of the dominant classical stochastic noise sources for $\mathrm{aLIGO} \; \mathrm{O3}$ (for an order of magnitude estimate), gives $\mathcal{S}_{\dot{N}}^\phi(\epsilon) \sim 6 \; \mathrm{photon} \;  \mathrm{s}^{-1} \cdot \mathrm{Hz}^{-1}$, at around $100 \; \mathrm{Hz}$ peak sensitivity, now assuming a filter cavity band-width of around $1 \; \mathrm{Hz}$ (about two orders of magnitude beyond state of the art from Ref.~\cite{PhysRevD.105.082003}, with approximately constant $\mathcal{S}_{\dot{N}}^\phi(\epsilon)$ in this bandwidth) - this would produce a noise-photon rate of $6 \; \mathrm{photon} \;  \mathrm{s}^{-1}$ (integrating an approximately constant noise flux PSD over this small bandwidth), much smaller than the $10^{6} \; \mathrm{photon} \;  \mathrm{s}^{-1}$ predicted in Ref.~\cite{6dzy-4c45} for the signal. These provide promising initial noise estimates for graviton absorption with energy read-out. 

Most importantly, as opposed to the resonant mass detector proposals, this interferometric version now corresponds to a gravitational wave detector that has detected known sources of gravitational radiation, while the Q-factor requirements for the resonant mass detector version of Ref.~\cite{TobarGermain2024Dsgw} are incredibly severe (just to obtain unity signal to noise (SNR) ratio requires optimised Q-factors and thermal noise that is extremely demanding). This motivates further investigation into the development of an interferometric graviton detector (in particular due to the very large signal to noise ratio implied by existing interferometric set-ups as considered here - an SNR of $10^5$). 

In essence the interferometric version, based on the initial quick analysis here using the translation of typical classical stochastic noise sources in the interferometric set-up - will beat the resonant mass detector version in signal to noise ratio by many orders of magnitude (even accounting for extremely optimised Q-factors for the resonant mass detector version with years of research and development). This suggests current experimental efforts should reconsider which approach to prioritise and motivates a more comprehensive analysis before committing substantial resources to either approach.

\section*{References}
\bibliography{moriond}

@article{TobarGermain2024Dsgw,
issn = {2041-1723},
journal = {Nature communications},
pages = {7229--14},
volume = {15},
publisher = {Nature Publishing Group UK},
number = {1},
year = {2024},
title = {Detecting single gravitons with quantum sensing},
copyright = {The Author(s) 2024},
language = {eng},
address = {London},
author = {Tobar, Germain and Manikandan, Sreenath K. and Beitel, Thomas and Pikovski, Igor},
organization = {Stevens Institute of Technology, Hoboken, NJ (United States)},
}

@misc{shenderov2024,
      title={Stimulated absorption of single gravitons: First light on quantum gravity, arXiv:2407.11929
}, 
      author={Victoria Shenderov and Mark Suppiah and Thomas Beitel and Germain Tobar and Sreenath K. Manikandan and Igor Pikovski},
      year={2024},
      eprint={2407.11929},
      archivePrefix={arXiv},
      primaryClass={gr-qc},
      url={https://arxiv.org/abs/2407.11929}, 
}

@article{PhysRevD.9.853,
  title = {Experimental distinction between the quantum and classical field-theoretic predictions for the photoelectric effect},
  author = {Clauser, John F.},
  journal = {Phys. Rev. D},
  volume = {9},
  issue = {4},
  pages = {853--860},
  numpages = {0},
  year = {1974},
  month = {Feb},
  publisher = {American Physical Society},
  doi = {10.1103/PhysRevD.9.853},
  url = {https://link.aps.org/doi/10.1103/PhysRevD.9.853}
}

@article{Boughn_2006,
doi = {10.1088/0264-9381/23/20/006},
url = {https://dx.doi.org/10.1088/0264-9381/23/20/006},
year = {2006},
month = {Sep},
publisher = {},
volume = {23},
number = {20},
pages = {5839},
author = {Boughn, Stephen and Rothman, Tony},
title = {Aspects of graviton detection: graviton emission and absorption by atomic hydrogen},
journal = {Classical and Quantum Gravity}
}

@article{TobarGermain2025Dkgf,
issn = {0264-9381},
journal = {Classical and Quantum Gravity},
volume = {42},
publisher = {IOP Publishing},
number = {5},
year = {2025},
title = {Detecting kHz gravitons from a neutron star merger with a multi-mode resonant mass detector},
copyright = {2025 The Author(s). Published by IOP Publishing Ltd},
language = {eng},
author = {Tobar, Germain and Pikovski, Igor and Tobar, Michael E},
}

@article{PhysRevLett.127.081602,
  title = {Quantum Mechanics of Gravitational Waves},
  author = {Parikh, Maulik and Wilczek, Frank and Zahariade, George},
  journal = {Phys. Rev. Lett.},
  volume = {127},
  issue = {8},
  pages = {081602},
  numpages = {5},
  year = {2021},
  month = {Aug},
  publisher = {American Physical Society},
  doi = {10.1103/PhysRevLett.127.081602},
  url = {https://link.aps.org/doi/10.1103/PhysRevLett.127.081602}
}

@article{parikh2020noise,
  title={The noise of gravitons},
  author={Parikh, Maulik and Wilczek, Frank and Zahariade, George},
  journal={International Journal of Modern Physics D},
  volume={29},
  number={14},
  pages={2042001},
  year={2020},
  publisher={World Scientific}
}

@article{PhysRevD.105.082003,
  title = {Improving the stability of frequency-dependent squeezing with bichromatic control of filter cavity length, alignment, and incident beam pointing},
  author = {Zhao, Yuhang and Capocasa, Eleonora and Eisenmann, Marc and Aritomi, Naoki and Page, Michael and Guo, Yuefan and Polini, Eleonora and Arai, Koji and Aso, Yoichi and van Beuzekom, Martin and Huang, Yao-Chin and Lee, Ray-Kuang and L{\"u}ck, Harald and Miyakawa, Osamu and Prat, Pierre and Shoda, Ayaka and Tacca, Matteo and Takahashi, Ryutaro and Vahlbruch, Henning and Vardaro, Marco and Wu, Chien-Ming and Leonardi, Matteo and Barsuglia, Matteo and Flaminio, Raffaele},
  journal = {Phys. Rev. D},
  volume = {105},
  issue = {8},
  pages = {082003},
  numpages = {11},
  year = {2022},
  month = {Apr},
  publisher = {American Physical Society},
  doi = {10.1103/PhysRevD.105.082003},
  url = {https://link.aps.org/doi/10.1103/PhysRevD.105.082003}
}

@article{PhysRevX.15.011034,
  title = {Photon-Counting Interferometry to Detect Geontropic Space-Time Fluctuations with GQuEST},
  author = {Vermeulen, Sander M. and Cullen, Torrey and Grass, Daniel and MacMillan, Ian A. O. and Ramirez, Alexander J. and Wack, Jeffrey and Korzh, Boris and Lee, Vincent S. H. and Zurek, Kathryn M. and Stoughton, Chris and McCuller, Lee},
  journal = {Phys. Rev. X},
  volume = {15},
  issue = {1},
  pages = {011034},
  numpages = {32},
  year = {2025},
  month = {Feb},
  publisher = {American Physical Society},
  doi = {10.1103/PhysRevX.15.011034},
  url = {https://link.aps.org/doi/10.1103/PhysRevX.15.011034}
}

@article{Tobar_2023_CSL,
year = {2023},
month = {Jul},
publisher = {IOP Publishing},
volume = {8},
number = {4},
pages = {045003},
author = {Germain Tobar and Stefan Forstner and Arkady Fedorov and Warwick P Bowen},
title = {Testing spontaneous wavefunction collapse with quantum electromechanics},
journal = {Quantum Science and Technology}
}

@article{carney2023,
  title = {Graviton detection and the quantization of gravity},
  author = {Carney, Daniel and Domcke, Valerie and Rodd, Nicholas L.},
  journal = {Phys. Rev. D},
  volume = {109},
  issue = {4},
  pages = {044009},
  numpages = {19},
  year = {2024},
  month = {Feb},
  publisher = {American Physical Society},
  doi = {10.1103/PhysRevD.109.044009},
  url = {https://link.aps.org/doi/10.1103/PhysRevD.109.044009}
}

@article{DYSONFREEMAN2013IAGD,
issn = {0217-751X},
journal = {International journal of modern physics. A, Particles and fields, gravitation, cosmology},
pages = {1330041},
volume = {28},
publisher = {World Scientific Publishing Company},
number = {25},
year = {2013},
title = {IS A GRAVITON DETECTABLE?},
copyright = {2013, World Scientific Publishing Company},
address = {Singapore},
author = {Dyson, Freeman},
}

@article{TOBAR2025139785,
title = {The gravito-phononic effect: a quantum signature of linearised gravity},
journal = {Physics Letters B},
volume = {869},
pages = {139785},
year = {2025},
issn = {0370-2693},
doi = {https://doi.org/10.1016/j.physletb.2025.139785},
url = {https://www.sciencedirect.com/science/article/pii/S0370269325005465},
author = {Germain Tobar and Oscar Berg}
}

@article{bose2017spin,
  title={Spin entanglement witness for quantum gravity},
  author={Bose, Sougato and Mazumdar, Anupam and Morley, Gavin W and Ulbricht, Hendrik and Toro{\v{s}}, Marko and Paternostro, Mauro and Geraci, Andrew A and Barker, Peter F and Kim, MS and Milburn, Gerard},
  journal={Physical Review Letters},
  volume={119},
  number={24},
  pages={240401},
  year={2017},
  publisher={APS}
}

@article{marletto2017gravitationally,
  title={Gravitationally induced entanglement between two massive particles is sufficient evidence of quantum effects in gravity},
  author={Marletto, Chiara and Vedral, Vlatko},
  journal={Physical Review Letters},
  volume={119},
  number={24},
  pages={240402},
  year={2017},
  publisher={APS}
}

@article{6dzy-4c45,
  title = {Wave-particle duality in the measurement of gravitational radiation},
  author = {Loughlin, Hudson A. and Tobar, Germain and Hall, Evan D. and Sudhir, Vivishek},
  journal = {Phys. Rev. Res.},
  volume = {7},
  issue = {4},
  pages = {043286},
  numpages = {19},
  year = {2025},
  month = {Dec},
  publisher = {American Physical Society},
  doi = {10.1103/6dzy-4c45},
  url = {https://link.aps.org/doi/10.1103/6dzy-4c45}
}

@article{Guerreiro2022quantumsignaturesin,
  title = {Quantum signatures in nonlinear gravitational waves},
  author = {Guerreiro, Thiago and Coradeschi, Francesco and Frassino, Antonia Micol and West, Jennifer Rittenhouse and Schioppa, Enrico Junior},
  journal = {{Quantum}},
  issn = {2521-327X},
  publisher = {{Verein zur F{\"{o}}rderung des Open Access Publizierens in den Quantenwissenschaften}},
  volume = {6},
  pages = {879},
  month = dec,
  year = {2022}
}

\end{document}